# Symmetries in Bosonization

**Yuan K. Ha**

Department of Physics, Temple University, Philadelphia, Pennsylvania 19122 U.S.A.

Email: yuanha@temple.edu                    Date: September 1, 2013

**Abstract**. Two-dimensional quantum field theories are important in many problems in physics because they contain exact symmetries and are often completely integrable. We demonstrate the power of bosonization in elucidating the structure of a multi-component fermion theory with chiral and internal symmeties. Many remarkable features and quantum symmetries in such a system are most readily uncovered in the bosonized form of the original fermion theory.

## 1. Introduction

In two-dimensional quantum field theories it is possible to express a fermion theory in terms of a boson theory to uncover hidden structure and new properties. This is known as bosonization. The equivalence has been demonstrated both at the Hamiltonian level and at the Lagrangian level for a number of theories. It is a remarkable occurrence in two-dimensional quantum field theory.

The goal of any symmetry operation is to leave the equations of a system invariant and therefore the physics unchanged. It is a search of self-similarity. Bosonization is a mathematical transformation that leaves the physics of a theory invariant even though the equations appear to be different. It is therefore a search of the self-similarity of the system and the operation can be considered as a quantum symmetry operation.

In two dimensions, Lagrangians have mass dimension 2. Fermions have dimension 1/2; bosons have dimension 0. Derivatives have dimension 1. When a massless fermion becomes a massive boson, the coupling constant in the fermion theory becomes the mass parameter in the boson theory. This is dynamical mass generation via bosonization. Many important symmetries in the fermion theory, such as chiral symmetry, internal symmetry, non-abelian symmetry, duality and triality, are immediately found in the bosonized version of the original theory.

## 2. Boson Representation

Let $\psi$ be a canonical Fermi field and $\phi$ a canonical Bose field in two dimensions. The boson representation of a free fermion in terms of a free boson is given by [1]

$$\psi(x) = \begin{bmatrix} \psi_+(x) \\ \psi_-(x) \end{bmatrix} = \frac{1}{(2\pi\alpha)^{1/2}} \begin{bmatrix} \exp[+i\sqrt{4\pi}\phi_+(x)] \\ \exp[-i\sqrt{4\pi}\phi_-(x)] \end{bmatrix}, \qquad (1)$$

where $\psi_+(x)$ and $\psi_-(x)$ are the chiral componets of the fermion. Each fermion component obeys the anti-commutation rules

$$\{\psi_\pm(x),\psi_\pm^\dagger(y)\} = \delta(x-y)$$
$$\{\psi_\pm(x),\psi_\mp^\dagger(y)\} = 0. \tag{2}$$

On the other hand, the components $\phi_+(x), \phi_-(x)$ are the positive and negative frequency parts of the scalar boson $\phi(x)$ satisfying the commutation rules

$$[\phi_\pm(x),\phi_\pm(y)] = \pm\varepsilon(x-y)$$
$$[\phi_+(x),\phi_-(y)] = i/4, \tag{3}$$

with $\phi_+(x)+\phi_-(x) = \phi(x)$, and $\varepsilon(x-y)$ is the odd step function with absolute value of unity. The quantity $1/\alpha$ is an ultraviolet cutoff imposed to guarantee the scalar field to remain finite at high frequencies, and $\alpha$ will be taken to zero at the end of any calculation involving the fermion $\psi(x)$. With the boson representation defined in Eq.(1), all the key properties of a quantum field theory such as correlation functions, cluster properties, Lorentz invariance are all found to be valid under this transformation. In addition, the Hamiltonian for the free fermion can be directly transformed into the Hamiltonian of the free boson. In this respect, we state the formal equivalence of bilinear forms appearing in the Lagrangians of the fermion theory and the boson theory in the following:

$$i\bar\psi\gamma^\mu\partial_\mu\psi \leftrightarrow \frac{1}{2}\partial_\mu\phi\partial^\mu\phi$$
$$\bar\psi\psi \leftrightarrow -\frac{1}{\pi\alpha}\cos\sqrt{4\pi}\phi$$
$$i\bar\psi\gamma^5\psi \leftrightarrow \frac{1}{\pi\alpha}\sin\sqrt{4\pi}\phi \tag{4}$$
$$\bar\psi\gamma^\mu\psi \leftrightarrow \frac{1}{\sqrt\pi}\varepsilon^{\mu\nu}\partial_\nu\phi$$
$$\bar\psi\gamma^5\gamma^\mu\psi \leftrightarrow \frac{1}{\sqrt\pi}\partial^\mu\phi.$$

As a demonstration of the instant power of bosonization, we can show directly that the massless Thirring model, a self-interacting fermion theory, is in fact equivalent to a free scalar boson theory [2]. The massless Thirring model, with coupling constant $g$, is given by the Lagrangian

$$L = i\bar\psi\gamma^\mu\partial_\mu\psi - \frac{g}{2}(\bar\psi\gamma^\mu\psi)(\bar\psi\gamma_\mu\psi). \tag{5}$$

The conservation of the fermion currents in this model shows that the currents can be expressed in terms of a new scalar field $\tilde\phi$. The requirement that this new field satisfies the canonical commutation relation for a scalar boson demands that it be rescaled from the free boson $\phi$ by the factor $\sqrt{1+g/\pi}$, i.e. $\tilde\phi = \phi/\sqrt{1+g/\pi}$. Applying the equivalence of the bilinear terms in the Thirring Lagrangian, we find that the fermion kinetic term produces a canonical boson kinetic term in terms of the new scalar field $\tilde\phi$, while the vector currents in the Lagrangian contribute an additional anomalous part to the boson kinetic term. The new scalar field and the contribution from the fermion interaction exactly compensate each other and the total effect of the bosonized Lagrangian is simply that of a free scalar boson given by

$$L = \left(1 + \frac{g}{\pi}\right) \times \frac{1}{2} \frac{\partial_\mu \phi}{\sqrt{1 + g/\pi}} \frac{\partial^\mu \phi}{\sqrt{1 + g/\pi}} = \frac{1}{2} \partial_\mu \phi \partial^\mu \phi. \tag{6}$$

If a mass term $m\bar{\psi}\psi$ is included in the Lagrangian of the massless Thirring model, the result is the massive Thirring model. It is well known that the massive Thirring model is equivalent to the quantum sine-Gordon theory [3]. The equivalence is most readily obtained by direct bosonization. The two Lagrangians are respectively,

$$L = i\bar{\psi}\gamma^\mu \partial_\mu \psi - \frac{g}{2}(\bar{\psi}\gamma^\mu \psi)(\bar{\psi}\gamma_\mu \psi) - m\bar{\psi}\psi$$
$$L = \frac{1}{2}\partial_\mu \phi \partial^\mu \phi + \frac{m}{\pi\alpha}\cos\sqrt{\frac{4\pi}{1+g/\pi}}\phi. \tag{7}$$

## 3. A multi-fermion theory

It is the purpose in this article to demonstrate that the real power of bosonization is in the elucidation of the structure of a multi-component fermion theory, in particular, the uncovering of hidden symmetries inaccessible by other means. Bosonization is uniquely a quantum field theory phenomenon; therefore any symmetry discovered which is not explicitly present in the fermion Lagrangian is a quantum symmetry. It is also shown that the number of fermion components can significantly alter the dynamics.

For our investigation, we consider a multi-component fermion theory with global non-abelian symmetry. This is the $U(N)$ Thirring model constructed with $N$ types of massless fermions $\psi^a$, $a = 1,...,N$ [4]. The interaction is through the fermion isospin currents. This model bears similarity to the Kondo problem [5] and is important in the study of many two-dimensional systems. It is of sufficient complexity that it cannot be solved easily. Bosonization can offer rare insights into the structure of this model. By a Lie algebraic generalization of the Fierz identity in two dimensions, the theory is seen to be equivalent to the chiral Gross-Neveu model [6]. The $U(N)$ Thirring Lagrangian is

$$L = \sum_{a=1}^{N} i\bar{\psi}^a \gamma^\mu \partial_\mu \psi^a - g \sum_{A=0}^{N^2-1}\left[\bar{\psi}\gamma^\mu \frac{\lambda^A}{2}\psi\right]\left[\bar{\psi}\gamma_\mu \frac{\lambda^A}{2}\psi\right], \tag{8}$$

where $g$ is the coupling constant and $\lambda^A$ are the generators of the $U(N)$ group. Upon standard bosonization, the theory becomes a system of $N$ coupled bosons $\phi^a$ with Lagrangian

$$L = \sum_{a=1}^{N} \frac{1}{2}\partial_\mu \phi^a \partial^\mu \phi^a + \sum_{a \neq b=1}^{N} \frac{g}{4\pi^2 \alpha^2}\cos\sqrt{\frac{4\pi}{1+g/\pi}}(\phi^a - \phi^b). \tag{9}$$

We may perform an orthogonal transformation $\Lambda$ on the $\phi^a$ fields in the following:

$$B^1 = (\phi^1 + \phi^2 + \ldots\ldots\ldots + \phi^N)/\sqrt{N}$$
$$B^2 = (\phi^1 - \phi^2)/\sqrt{2}$$
$$B^3 = (\phi^1 + \phi^2 - 2\phi^3)/\sqrt{6} \qquad (10)$$
$$\ldots\ldots$$
$$\ldots\ldots$$
$$B^N = (\phi^1 + \ldots\ldots + \phi^{N-1} - (N-1)\phi^N)/\sqrt{N(N-1)}.$$

In this case, a massless free boson $B^1$ will decouple and the effective theory becomes a system of $N-1$ coupled sine-Gordon theories with Lagrangian

$$L = \frac{1}{2}\partial_\mu B^1 \partial^\mu B^1 + \sum_{k=2}^{N}\frac{1}{2}\partial_\mu B^k \partial^\mu B^k + \sum_{i\neq j=2}^{N}\frac{g}{4\pi^2\alpha^2}\cos\sqrt{\frac{4\pi}{1+g/\pi}}(\Lambda^{li}B^l - \Lambda^{mj}B^m). \qquad (11)$$

This is a realization of the symmetry $U(N) \approx SU(N) \otimes U(1)$ in the boson theory. Some remarkable equivalences can now be found from the interacting boson Lagrangian from Eq.(11).

3.1. $N=2$ *Case*. The Lagrangian is that of a free boson and a sine-Gordon theory with symmetry $U(2) \approx SU(2) \otimes U(1)$:

$$L = \frac{1}{2}\partial_\mu B^1 \partial^\mu B^1 + \frac{1}{2}\partial_\mu B^2 \partial^\mu B^2 + \frac{g}{4\pi^2\alpha^2}\cos\sqrt{\frac{8\pi}{1+g/\pi}}B^2. \qquad (12)$$

If we take two boson Lagrangians with identical couplings from Eq.(9), one with bosons $(\phi^1, \phi^2)$ and the other with bosons $(n^1, n^2)$, we find the following situation:

$$L = \left(\frac{1}{2}\partial_\mu \phi^1 \partial^\mu \phi^1 + \frac{1}{2}\partial_\mu \phi^2 \partial^\mu \phi^2\right) + \left(\frac{1}{2}\partial_\mu n^1 \partial^\mu n^1 + \frac{1}{2}\partial_\mu n^2 \partial^\mu n^2\right)$$
$$+ \frac{g}{4\pi^2\alpha^2}\cos\sqrt{\frac{4\pi}{1+g/\pi}}(\phi^1 - \phi^2) + \frac{g}{4\pi^2\alpha^2}\cos\sqrt{\frac{4\pi}{1+g/\pi}}(n^1 - n^2). \qquad (13)$$

This is a system of two independent $U(2)$ models with four bosons. If we further allow the four bosons to mix according to the following orthogonal transformation,

$$B^1 = \frac{1}{2}(\phi^1 + \phi^2 + n^1 + n^2)$$

$$B^2 = \frac{1}{2}(\phi^1 + \phi^2 - n^1 - n^2)$$

$$B^3 = \frac{1}{2}(\phi^1 - \phi^2 + n^1 - n^2) \quad (14)$$

$$B^4 = \frac{1}{2}(\phi^1 - \phi^2 - n^1 + n^2),$$

we find the fields $(\phi^1 - \phi^2) = (B^3 + B^4)$ and $(n^1 - n^2) = (B^3 - B^4)$. The system now becomes two coupled sine-Gordon theories with two interacting bosons $(B^3, B^4)$, together with two massless free bosons $(B^1, B^2)$. The ensuing Lagrangian then becomes

$$L = \frac{1}{2}\partial_\mu B^1 \partial^\mu B^1 + \frac{1}{2}\partial_\mu B^2 \partial^\mu B^2 + \frac{1}{2}\partial_\mu B^3 \partial^\mu B^3 + \frac{1}{2}\partial_\mu B^4 \partial_\mu B^4$$

$$+ \frac{g}{4\pi^2 \alpha^2} \cos\sqrt{\frac{4\pi}{1+g/\pi}}(B^3 + B^4) + \frac{g}{4\pi^2 \alpha^2} \cos\sqrt{\frac{4\pi}{1+g/\pi}}(B^3 - B^4). \quad (15)$$

At this stage, it is auspicious to note that the trigonometric identity

$$\cos(A+B) + \cos(A-B) = 2\cos A \cos B \quad (16)$$

is exclusively applicable to Bose fields in two dimensions. The Lagrangian can be transformed into

$$L = \frac{1}{2}\partial_\mu B^1 \partial^\mu B^1 + \frac{1}{2}\partial_\mu B^2 \partial^\mu B^2 + \frac{1}{2}\partial_\mu B^3 \partial^\mu B^3 + \frac{1}{2}\partial_\mu B^4 \partial_\mu B^4$$

$$+ \frac{g}{2\pi^2 \alpha^2} \cos\sqrt{\frac{4\pi}{1+g/\pi}} B^3 \times \cos\sqrt{\frac{4\pi}{1+g/\pi}} B^4. \quad (17)$$

The interacting part with Bose fields $(B^3, B^4)$ is remarkably the bosonized form of the $O(4)$ Gross-Neveu model [7] constructed with four Majorana fermions $(\chi^1, \chi^2, \chi^3, \chi^4)$, the Lagrangian for which is

$$L = i\bar{\chi}^1 \gamma^\mu \partial_\mu \chi^1 + i\bar{\chi}^2 \gamma^\mu \partial_\mu \chi^2 + i\bar{\chi}^3 \gamma^\mu \partial_\mu \chi^3 + i\bar{\chi}^4 \gamma^\mu \partial_\mu \chi^4$$

$$+ \frac{g}{2}\left[\bar{\chi}^1 \chi^1 + \bar{\chi}^2 \chi^2 + \bar{\chi}^3 \chi^3 + \bar{\chi}^4 \chi^4\right]^2. \quad (18)$$

This is a realization of the symmetry $SU(2) \otimes SU(2) \approx O(4)$ through two different fermion theories. The sine-Gordon Lagrangian has a hidden $SU(2)$ symmetry.

3.2. *N* = 4 *Case.* The bosonized Lagrangian from the $U(4)$ Thirring model in Eq.(9) takes the form

$$L = \frac{1}{2}\partial_\mu\phi^1\partial^\mu\phi^1 + \frac{1}{2}\partial_\mu\phi^2\partial^\mu\phi^2 + \frac{1}{2}\partial_\mu\phi^3\partial^\mu\phi^3 + \frac{1}{2}\partial_\mu\phi^4\partial^\mu\phi^4$$

$$+ \frac{g}{4\pi^2\alpha^2}\left(\begin{array}{l}\cos\sqrt{\frac{4\pi}{1+g/\pi}}(\phi^1-\phi^2)+\cos\sqrt{\frac{4\pi}{1+g/\pi}}(\phi^3-\phi^4)+\\ \cos\sqrt{\frac{4\pi}{1+g/\pi}}(\phi^1-\phi^3)+\cos\sqrt{\frac{4\pi}{1+g/\pi}}(\phi^2-\phi^4)+\\ \cos\sqrt{\frac{4\pi}{1+g/\pi}}(\phi^1-\phi^4)+\cos\sqrt{\frac{4\pi}{1+g/\pi}}(\phi^2-\phi^3)\end{array}\right). \quad (19)$$

We may again decouple a massless boson $B^1$ by applying the following transformation

$$\begin{aligned} B^1 &= \frac{1}{2}(\phi^1+\phi^2+\phi^3+\phi^4) \\ B^2 &= \frac{1}{2}(\phi^1+\phi^2-\phi^3-\phi^4) \\ B^3 &= \frac{1}{2}(\phi^1-\phi^2+\phi^3-\phi^4) \\ B^4 &= \frac{1}{2}(\phi^1-\phi^2-\phi^3+\phi^4). \end{aligned} \quad (20)$$

There are six interacting terms in Eq.(19) in the form of cosine fields but after the transformation they may be combined using the previous trigonometric identity into three product terms in the following form:

$$L = \frac{1}{2}\partial_\mu B^1\partial^\mu B^1 + \frac{1}{2}\partial_\mu B^2\partial^\mu B^2 + \frac{1}{2}\partial_\mu B^3\partial^\mu B^3 + \frac{1}{2}\partial_\mu B^4\partial^\mu B^4$$

$$+ \frac{g}{2\pi^2\alpha^2}\left(\begin{array}{l}\cos\sqrt{\frac{4\pi}{1+g/\pi}}B^2 \times \cos\sqrt{\frac{4\pi}{1+g/\pi}}B^3 + \\ \cos\sqrt{\frac{4\pi}{1+g/\pi}}B^3 \times \cos\sqrt{\frac{4\pi}{1+g/\pi}}B^4 + \\ \cos\sqrt{\frac{4\pi}{1+g/\pi}}B^4 \times \cos\sqrt{\frac{4\pi}{1+g/\pi}}B^2\end{array}\right). \quad (21)$$

The effective theory with three bosons $(B^2, B^3, B^4)$ is remarkably the bosonized form of the $O(6)$ Gross-Neveu model with six Majorano fermions $(\chi^1, \chi^2, \chi^3, \chi^4, \chi^5, \chi^6)$, the Lagrangian for which explicitly is

$$L = i\bar{\chi}^1 \gamma^\mu \partial_\mu \chi^1 + i\bar{\chi}^2 \gamma^\mu \partial_\mu \chi^2 + i\bar{\chi}^3 \gamma^\mu \partial_\mu \chi^3 + i\bar{\chi}^4 \gamma^\mu \partial_\mu \chi^4 + i\bar{\chi}^5 \gamma^\mu \partial_\mu \chi^5 + i\bar{\chi}^6 \gamma^\mu \partial_\mu \chi^6$$
$$+ \frac{g}{2}\left[\bar{\chi}^1 \chi^1 + \bar{\chi}^2 \chi^2 + \bar{\chi}^3 \chi^3 + \bar{\chi}^4 \chi^4 + \bar{\chi}^5 \chi^5 + \bar{\chi}^6 \chi^6\right]^2. \tag{22}$$

This is a surprising realization of the symmetry $SU(4) \approx O(6)$ through the fermion theories.

**4. A novel symmetry**

We return once again to the original bosonized Lagrangian for the $U(N)$ Thirring model in Eq.(9) and this time apply a novel transformation for all cases with $N \geq 3$. This will allow us to uncover a hidden structure which is not evident in the fermion theory. The smallest number for $N$ here is 3 since one of the bosons will later decouple and at least two remaining bosons are necessary in order to form a new linear combination. The transformation is given through the following $N \times N$ matrix,

$$\begin{pmatrix} B^1 \\ B^2 \\ . \\ . \\ . \\ . \\ B^{N-1} \\ B^N \end{pmatrix} = \frac{1}{N} \begin{pmatrix} 2 & 2 & \ldots & 2 & 2-N \\ 2 & 2 & \ldots & 2-N & 2 \\ . & . & \ldots & . & . \\ . & . & \ldots & . & . \\ . & . & \ldots & . & . \\ . & . & \ldots & . & . \\ 2 & 2-N & \ldots & 2 & 2 \\ 2-N & 2 & \ldots & 2 & 2 \end{pmatrix} \begin{pmatrix} \phi^1 \\ \phi^2 \\ . \\ . \\ . \\ . \\ \phi^{N-1} \\ \phi^N \end{pmatrix}. \tag{23}$$

The Bose fields $(\phi^1, \ldots, \phi^N)$ are now replaced by the set $(B^1, \ldots, B^N)$. In the above matrix, all the elements have a value equal to $2$, except one element in each row and each column has a value equal to $2-N$. The most remarkable property of this transformation is that the difference of any two $\phi$ fields becomes the difference of two $B$ fields. The total number of possible field differences is the same in both sets. This transformation generates a new symmetry known as boson self-duality. There are two identical boson Lagrangians:

$$L = \sum_{a=1}^{N} \frac{1}{2} \partial_\mu \phi^a \partial_\mu \phi^a + \sum_{a \neq b=1}^{N} \frac{g}{4\pi^2 \alpha^2} \cos\sqrt{\frac{4\pi}{1+g/\pi}}(\phi^a - \phi^b)$$
$$\tag{24}$$
$$L = \sum_{a=1}^{N} \frac{1}{2} \partial_\mu B^a \partial_\mu B^a + \sum_{a \neq b=1}^{N} \frac{g}{4\pi^2 \alpha^2} \cos\sqrt{\frac{4\pi}{1+g/\pi}}(B^a - B^b).$$

They correspond to two identical fermion Lagrangians which display fermion self-duality. In terms of the equivalent chiral Gross-Neveu model, they are respectively,

$$L = \sum_{a=1}^{N} i\bar{\psi}^a \gamma^\mu \partial_\mu \psi^a + \frac{g}{2}\left\{[\sum_{a=1}^{N} \bar{\psi}^a \psi^a]^2 - [\sum_{a=1}^{N} \bar{\psi}^a \gamma^5 \psi^a]^2\right\}$$

$$L = \sum_{a=1}^{N} i\bar{\xi}^a \gamma^\mu \partial_\mu \xi^a + \frac{g}{2}\left\{[\sum_{a=1}^{N} \bar{\xi}^a \xi^a]^2 - [\sum_{a=1}^{N} \bar{\xi}^a \gamma^5 \xi^a]^2\right\}.$$

(25)

The Fermi fields $(\psi^1,...,\psi^N)$ are replaced by the dual set $(\xi^1,...,\xi^N)$ through bosonization. In self-duality, each degree of freedom in one form is a linear or non-linear combination of all the degrees of freedom in the other form. Symmetry dictates interaction. The presence of self-duality symmetry in the dynamics of the $U(N)$ Thirring model is to maintain the equality of masses of the kinks and the particles in the quantum theory. We also see self-triality symmetry in the special case of the O(8) Gross-Neveu model [8]. There are three identical fermion Lagrangians which are connected with one another through their bosons. Duality and triality are discrete symmetries but they are not related to any conservation law. They are important quantum symmetries which are most elegantly found in bosonization.